\documentclass[10pt]{article}
\usepackage{graphicx}
\usepackage{amsmath}
\usepackage{amssymb}
\usepackage{caption2}
\begin{document}
\begin{center}
{\large {\bf \sc{  Scalar or  vector tetraquark state candidate: $Z_c(4100)$  }}} \\[2mm]
Zhi-Gang  Wang \footnote{E-mail: zgwang@aliyun.com.  }     \\
 Department of Physics, North China Electric Power University, Baoding 071003, P. R. China
\end{center}

\begin{abstract}
In this article,  we separate the vector and axialvector components of the tensor diquark operators explicitly, construct the axialvector-axialvector type  and vector-vector type scalar tetraquark currents and scalar-tensor type tensor tetraquark current to study the scalar, vector and axialvector tetraquark states with the QCD sum rules in a consistent way.  The present calculations do not favor assigning the $Z_c(4100)$ to be a scalar or vector tetraquark state. If the $Z_c(4100)$ is a scalar tetraquark state without mixing effects, it should have a mass about $3.9\,\rm{GeV}$ or $4.0\,\rm{GeV}$ rather than $4.1\,\rm{GeV}$; on the other hand, if the $Z_c(4100)$ is a vector tetraquark state, it should have a mass about  $4.2\,\rm{GeV}$ rather than $4.1\,\rm{GeV}$. However, if we introduce mixing,  a mixing scalar tetraquark state can  have a mass about $4.1\,\rm{GeV}$.  As a byproduct, we obtain an axialvector tetraquark candidate for the $Z_c(4020)$.
 \end{abstract}



\section{Introduction}
The attractive interactions of one-gluon exchange  favor  formation of
the diquarks in  color antitriplet  \cite{One-gluon}. The diquarks (or diquark operators) $\varepsilon^{abc}q^{T}_b C\Gamma q^{\prime}_c$ in  color antitriplet have  five  structures  in Dirac spinor space, where $C\Gamma=C$, $C\gamma_5$, $C\gamma_\mu \gamma_5$,  $C\gamma_\mu $ and $C\sigma_{\mu\nu}$ for the  pseudoscalar ($P$), scalar ($S$), vector ($V$), axialvector ($A$)  and  tensor ($T$) diquarks, respectively, the $a$, $b$, $c$ are color indexes. The  pseudoscalar, scalar, vector and axialvector  diquarks have been studied with the QCD sum rules in details, which indicate that  the favored configurations are the scalar  and axialvector  diquark states \cite{Dosch-Diquark-1989,WangDiquark,WangLDiquark}. The scalar  and axialvector  diquark operators  have been applied extensively to construct the tetraquark currents to study the lowest tetraquark states \cite{Nielsen-X3872,Wang-Scalar-1GeV, Wang-Axial-1GeV,WangHuangtao-2014-PRD, Wang-3915-CgmCgm, Wang-3915-C5C5, Wang-3860,Azizi-Zc4100}.

In 2018, the LHCb Collaboration performed a Dalitz plot analysis of  ${{B} ^0} \!\rightarrow \eta _c {{K} ^+} {{\pi } ^-}$ decays and observed an evidence for an exotic   $\eta _c {{\pi } ^-}$  resonant state \cite{LHCb-Z4100}.   The significance of this exotic resonance is more than three standard deviations, the measured  mass and width are $4096\pm 20  {}^{+18}_{-22}\,\rm{ MeV}$  and $152\pm 58 {}^{+60}_{-35}\,\rm{ MeV}$, respectively. The spin-parity assignments $J^P =0^+$ and $1^-$ are both consistent with the experimental data \cite{LHCb-Z4100}. Is it a candidate for a scalar tetraquark state \cite{Azizi-Zc4100,Wang-Vector-Zc4100,Wu-Zc4100}, hadrocharmonium \cite{Voloshin-Zc4100}, $D^*\bar{D}^*$ molecular state \cite{ZhaoQ-Zc4100}, or charge conjugation of the $Z^+_c(4050)$ \cite{Dai-Zc4100}\,?

In Refs.\cite{Wang-3915-CgmCgm,Wang-3915-C5C5}, we study the $[sc]_S[\bar{s}\bar{c}]_S$-type, $[sc]_A[\bar{s}\bar{c}]_A$-type,
$[sc]_P[\bar{s}\bar{c}]_P$-type and $[sc]_V[\bar{s}\bar{c}]_V$-type scalar tetraquark states with the QCD sum rules  systematically, and obtain the ground state masses $M_{SS}=3.89\pm0.05\,\rm{GeV}$ $(3.85^{+0.18}_{-0.17}\,\rm{GeV})$, $M_{AA}=3.92^{+0.19}_{-0.18}\,\rm{GeV}$, $M_{PP} = 5.48 \pm 0.10\,\rm{GeV}$ and $M_{VV}=4.70^{+0.08}_{-0.09}\,\rm{ GeV}$  for the $SS$, $AA$, $PP$ and $VV$ diquark-antidiquark type tetraquark states, respectively. The larger uncertainties ${}^{+0.18}_{-0.17}\,\rm{GeV}$ and ${}^{+0.19}_{-0.18}\,\rm{GeV}$ (compared to the uncertainties $\pm0.05\,\rm{GeV}$,  $\pm 0.10\,\rm{GeV}$ and ${}^{+0.08}_{-0.09}\,\rm{ GeV}$) in the ground state masses originate from the QCD sum rules, where  both the ground state and the first radial excited state are taken into account at the hadron side. If only the ground states are taken into account in the QCD sum rules, the uncertainties  $|\delta M|\leq 0.10\,\rm{GeV}$, so the uncertainties ${}^{+0.18}_{-0.17}\,\rm{GeV}$ and ${}^{+0.19}_{-0.18}\,\rm{GeV}$ ($\pm 0.10\,\rm{GeV}$) can be referred to as "larger uncertainties" ("smaller  uncertainties").

In Ref.\cite{Wang-3860}, we tentatively assign the $X^*(3860)$ to be the $[qc]_S[\bar{q}\bar{c}]_S$-type  scalar
tetraquark state, study its mass and width with the QCD sum rules in details, and obtain the mass $M_{SS} = 3.86 \pm 0.09\,\rm{GeV}$.
 Now we can see that  the $SU(3)$ breaking effects of the masses of the $[sc][\bar{s}\bar{c}]$ and $[qc][\bar{q}\bar{c}]$  tetraquark states from the QCD sum rules are rather  small, roughly speaking, they have degenerate  masses. If we take the larger uncertainty,  the predicted mass $M_{AA}=3.92^{+0.19}_{-0.18}\,\rm{GeV}$ for the  $[uc][\bar{d}\bar{c}]$  tetraquark state has overlap with the experimental data  $M_{Z_c}=4096 \pm 20^{+18}_{-22}\,\rm{MeV}$ marginally  \cite{LHCb-Z4100}, and favors assigning the $Z_c(4100)$ to be  the $AA$-type scalar tetraquark state \cite{Wang-Vector-Zc4100}. On the other hand, if we take the smaller uncertainty, the predicted mass $M_{AA}=3.92 \pm 0.10\,\rm{GeV}$ for the   $[uc][\bar{d}\bar{c}]$  tetraquark state has no overlap with the experimental data $M_{Z_c}=4096 \pm 20^{+18}_{-22}\,\rm{MeV}$, and disfavors assigning the $Z_c(4100)$ to be  the $AA$-type scalar tetraquark state. In a word, it is not robust assigning the $Z_c(4100)$ to be  the $AA$-type scalar tetraquark state.

In Ref.\cite{Azizi-Zc4100}, Sundu, Agaev and Azizi obtain the mass $M_{SS}=4.08 \pm 0.15\,\rm{GeV}$, which differs from the prediction $M_{SS} = 3.86 \pm 0.09\,\rm{GeV}$ greatly \cite{Wang-3860}. The differences  originate from the different input parameters at the QCD side and different pole contributions at the hadron side. In Ref.\cite{Wang-3860}, the pole contribution is about $(46-70)\%$, which is much larger than the pole contribution in Ref.\cite{Azizi-Zc4100}. While in early works, we took the pole contributions $>50\%$ and chose  the energy scales  of the QCD spectral densities to be $\mu=1\,\rm{GeV}$, and obtained almost  degenerate masses $M_{SS}\approx M_{AA}\approx 4.4\,\rm{GeV}$ \cite{Wang-Scalar-1GeV}, which are  much larger than the value $M_{SS}=4.08 \pm 0.15\,\rm{GeV}$ \cite{Azizi-Zc4100}. In Ref.\cite{WangHuangtao-2014-PRD}, we observe that the energy scale $\mu=1\,\rm{GeV}$ is not the optimal energy scale for the hidden-charm tetraquark states.

In summary, the QCD sum rules  do not favor assigning the $Z_c(4100)$ to be the $[qc]_S[\bar{q}\bar{c}]_S$-type, $[qc]_A[\bar{q}\bar{c}]_A$-type,
$[qc]_P[\bar{q}\bar{c}]_P$-type and $[qc]_V[\bar{q}\bar{c}]_V$-type scalar tetraquark states.

In Refs.\cite{Wang-Vector-Zc4100,Wang-Vector-tetraquark}, we take the scalar and axialvector diquark operators as basic constituents,
introduce an explicit P-wave between the diquark and antidiquark operators  to construct the vector tetraquark currents,
 and study the  vector tetraquark states with the QCD sum rules systematically, and obtain the lowest vector tetraquark masses up to now,
 $M_{Y}=4.24\pm0.10\,\rm{GeV}$, $4.28\pm0.10\,\rm{GeV}$, $4.31\pm0.10\,\rm{GeV}$  and $4.33\pm0.10\,\rm{GeV}$ for the tetraquark states $|0, 0; 0, 1; 1\rangle$, $|1, 1; 0, 1; 1\rangle$ , $\frac{1}{\sqrt{2}}\left(|1, 0; 1, 1; 1\rangle+|0, 1; 1, 1; 1\rangle\right)$  and $|1, 1; 2, 1; 1\rangle$,  respectively,   where the tetraquark states are defined by $|S_{[qc]}, S_{[\bar{q}\bar{c}]}; S, L; J\rangle$, the $S$, $L$ and $J$ are the diquark spin, angular momentum and total angular momentum,  respectively.
 For other QCD sum rules with the $[qc]_S\partial_\mu [\bar{q}\bar{c}]_S$-type interpolating  currents, one can consult Ref.\cite{ZhangJR-Y4660}.
 In fact, if we take the pseudoscalar and vector diquark operators as the basic constituents, even (or much) larger tetraquark masses are obtained \cite{Other-Y4660,WangEPJC-1601,Wang-tetra-formula}.
Up to now, it is obvious that the QCD sum rules do not favor assigning the $Z_c(4100)$ to be the vector tetraquark state.

It is interesting to analyze the properties of the tensor diquark states, and take  the tensor diquark operators as the basic constituents to construct the tetraquark currents  to study the $Z_c(4100)$.

 Under parity transform $\widehat{P}$, the tensor diquark  operators have the  properties,
\begin{eqnarray}
\widehat{P}\varepsilon^{abc}q^{Tb}(x)C\sigma_{\mu\nu}\gamma_5Q^c(x)\widehat{P}^{-1}&=&\varepsilon^{abc}q^{Tb}(\tilde{x}) C\gamma^0\sigma_{\mu\nu}\gamma^0\gamma_5Q^c(\tilde{x})\nonumber\\
&=&\varepsilon^{abc}q^{Tb}(\tilde{x})C\sigma^{\mu\nu}\gamma_5Q^c(\tilde{x}) \, , \nonumber\\
\widehat{P}\varepsilon^{abc}q^{Tb}(x)C\sigma_{\mu\nu} Q^c(x)\widehat{P}^{-1}&=&-\varepsilon^{abc}q^{Tb}(\tilde{x})C\gamma^0\sigma_{\mu\nu}\gamma^0 Q^c(\tilde{x})\nonumber\\
&=&-\varepsilon^{abc}q^{Tb}(\tilde{x})C\sigma^{\mu\nu} Q^c(\tilde{x}) \, ,
\end{eqnarray}
where the four vectors $x^\mu=(t,\vec{x})$ and $\tilde{x}^\mu=(t,-\vec{x})$, and we have used the properties of the Dirac $\gamma$-matrixes, $\gamma^0 \gamma_\mu\gamma^0=\gamma^\mu$ and $\gamma^0\sigma_{\mu\nu}\gamma^0=\sigma^{\mu\nu}$.   The tensor diquark states have both $J^P=1^+$ and $1^-$ components,
\begin{eqnarray}
\widehat{P}\varepsilon^{abc}q^{Tb}(x)C\sigma_{jk}\gamma_5Q^c(x)\widehat{P}^{-1}&=&+\varepsilon^{abc}q^{Tb}(\tilde{x})C\sigma_{jk}\gamma_5Q^c(\tilde{x}) \, , \nonumber\\
\widehat{P}\varepsilon^{abc}q^{Tb}(x)C\sigma_{0j}\gamma_5Q^c(x)\widehat{P}^{-1}&=&-\varepsilon^{abc}q^{Tb}(\tilde{x})C\sigma_{0j}\gamma_5Q^c(\tilde{x}) \, , \nonumber\\
\widehat{P}\varepsilon^{abc}q^{Tb}(x)C\sigma_{0j} Q^c(x)\widehat{P}^{-1}&=&+\varepsilon^{abc}q^{Tb}(\tilde{x})C\sigma_{0j} Q^c(\tilde{x}) \, , \nonumber\\
\widehat{P}\varepsilon^{abc}q^{Tb}(x)C\sigma_{jk} Q^c(x)\widehat{P}^{-1}&=&-\varepsilon^{abc}q^{Tb}(\tilde{x})C\sigma_{jk} Q^c(\tilde{x})\, ,
\end{eqnarray}
where $j$, $k=1$, $2$, $3$,  and we have used the properties of the Dirac $\gamma$-matrixes, $\gamma^0=\gamma_0$, $\gamma^j=-\gamma_j$, $\sigma^{jk}=\sigma_{jk}$ and $\sigma^{0j}=-\sigma_{0j}$.
Now we introduce the four vector $t^\mu=(1,\vec{0})$ and project out the $1^+$ and $1^-$ components explicitly,
\begin{eqnarray}
\widehat{P}\varepsilon^{abc}q^{Tb}(x)C\sigma^t_{\mu\nu}\gamma_5Q^c(x)\widehat{P}^{-1}&=&+\varepsilon^{abc}q^{Tb}(\tilde{x})C\sigma^t_{\mu\nu}\gamma_5Q^c(\tilde{x}) \, , \nonumber\\
\widehat{P}\varepsilon^{abc}q^{Tb}(x)C\sigma^v_{\mu\nu}\gamma_5Q^c(x)\widehat{P}^{-1}&=&-\varepsilon^{abc}q^{Tb}(\tilde{x})C\sigma^v_{\mu\nu}\gamma_5Q^c(\tilde{x}) \, ,\nonumber\\
\widehat{P}\varepsilon^{abc}q^{Tb}(x)C\sigma^v_{\mu\nu} Q^c(x)\widehat{P}^{-1}&=&+\varepsilon^{abc}q^{Tb}(\tilde{x})C\sigma^v_{\mu\nu} Q^c(\tilde{x}) \, , \nonumber\\
\widehat{P}\varepsilon^{abc}q^{Tb}(x)C\sigma^t_{\mu\nu} Q^c(x)\widehat{P}^{-1}&=&-\varepsilon^{abc}q^{Tb}(\tilde{x})C\sigma^t_{\mu\nu} Q^c(\tilde{x}) \, , 
\end{eqnarray}
where
\begin{eqnarray}
\sigma^t_{\mu\nu} &=&\frac{i}{2}\Big[\gamma^t_\mu, \gamma^t_\nu \Big]\, ,\nonumber \\
\sigma^v_{\mu\nu} &=&\frac{i}{2}\Big[\gamma^v_\mu, \gamma^t_\nu \Big]\, ,\nonumber \\
\gamma^v_\mu &=&  \gamma \cdot t t_\mu\, ,\nonumber \\
\gamma^t_\mu&=&\gamma_\mu-\gamma \cdot t t_\mu\, .
\end{eqnarray}
In this article, we choose the axialvector diquark operator $\varepsilon^{abc}q^{Tb}(x)C\sigma^v_{\mu\nu} Q^c(x)$ and vector diquark operator $\varepsilon^{abc}q^{Tb}(x)C\sigma^t_{\mu\nu} Q^c(x)$   to construct the tetraquark operators. On the other hand, if we choose the axialvector diquark operator $\varepsilon^{abc}q^{Tb}(x)C\sigma^t_{\mu\nu} \gamma_5Q^c(x)$ and vector diquark operator $\varepsilon^{abc}q^{Tb}(x)C\sigma^v_{\mu\nu} \gamma_5Q^c(x)$ to construct the tetraquark operators, we can obtain the same predictions.

In this article, we separate the vector ($\tilde{V}$) and axialvector ($\tilde{A}$) components of the tensor diquark operators explicitly, construct the $\tilde{A}\tilde{A}$-type  and $\tilde{V}\tilde{V}$-type scalar tetraquark currents, and the $ST$-type tensor tetraquark current to study the $Z_c(4100)$ with the QCD sum rules, and try to  assign  the $Z_c(4100)$ in the  scenario of scalar and vector tetraquark states once more.

The article is arranged as follows:  we derive the QCD sum rules for the masses and pole residues  of  the   tetraquark states in section 2; in section 3, we   present the numerical results and discussions; section 4 is reserved for our conclusion.

\section{QCD sum rules for  the  scalar and vector tetraquark states}
Firstly, we write down  the two-point correlation functions   in the QCD sum rules,
\begin{eqnarray}
\Pi(p)&=&i\int d^4x e^{ip \cdot x} \langle0|T\Big\{J(x)J^{\dagger}(0)\Big\}|0\rangle \, , \nonumber\\
\Pi_{\mu\nu\alpha\beta}(p)&=&i\int d^4x e^{ip \cdot x} \langle0|T\Big\{J_{\mu\nu}(x)J_{\alpha\beta}^{\dagger}(0)\Big\}|0\rangle \, ,
\end{eqnarray}
where $J(x)=J_{\tilde{A}\tilde{A}}(x)$, $J_{\tilde{V}\tilde{V}}(x)$,
\begin{eqnarray}
J_{\tilde{A}\tilde{A}}(x)&=&\varepsilon^{ijk}\varepsilon^{imn}u^{Tj}(x)C\sigma^v_{\mu\nu} c^k(x)  \bar{d}^m(x)\sigma_v^{\mu\nu} C \bar{c}^{Tn}(x) \, ,\nonumber \\
J_{\tilde{V}\tilde{V}}(x)&=&\varepsilon^{ijk}\varepsilon^{imn}u^{Tj}(x)C\sigma^t_{\mu\nu} c^k(x)  \bar{d}^m(x)\sigma_t^{\mu\nu} C \bar{c}^{Tn}(x) \, ,\nonumber \\
J_{\mu\nu}(x)&=&\frac{\varepsilon^{ijk}\varepsilon^{imn}}{\sqrt{2}}\Big[u^{Tj}(x)C\gamma_5 c^k(x)  \bar{d}^m(x)\sigma_{\mu\nu} C \bar{c}^{Tn}(x)- u^{Tj}(x)C\sigma_{\mu\nu} c^k(x)  \bar{d}^m(x)\gamma_5 C \bar{c}^{Tn}(x) \Big] \, , \nonumber\\
\end{eqnarray}
 the $i$, $j$, $k$, $m$, $n$ are color indexes. We take the isospin limit by assuming the $u$ and $d$ quarks have  degenerate masses.
 Under charge conjugation transform $\widehat{C}$, the currents $J(x)$ and $J_{\mu\nu}(x)$ have the properties,
\begin{eqnarray}
\widehat{C}J(x)\widehat{C}^{-1}&=&+ J(x) \, , \nonumber\\
\widehat{C}J_{\mu\nu}(x)\widehat{C}^{-1}&=&- J_{\mu\nu}(x) \, ,
\end{eqnarray}
 the currents have definite charge conjugation. The currents $J(x)$ couple potentially to the scalar tetraquark states, while the
  current $J_{\mu\nu}(x)$ couples potentially to both the  $J^{PC}=1^{+-}$ and $1^{--}$ tetraquark states,
\begin{eqnarray}
 \langle 0|J(0)|Z_c^+(p)\rangle &=&\lambda_{Z^+}\, , \nonumber\\
  \langle 0|J_{\mu\nu}(0)|Z_c^-(p)\rangle &=& \frac{\lambda_{Z^-}}{M_{Z^-}} \, \varepsilon_{\mu\nu\alpha\beta} \, \varepsilon^{\alpha}p^{\beta}\, , \nonumber\\
 \langle 0|J_{\mu\nu}(0)|Z_c^+(p)\rangle &=&\frac{\lambda_{Z^+}}{M_{Z^+}} \left(\varepsilon_{\mu}p_{\nu}-\varepsilon_{\nu}p_{\mu} \right)\, ,
\end{eqnarray}
the  $\varepsilon_{\mu/\alpha}$ are the polarization vectors of the  tetraquark states, the superscripts ${}^\pm$ denote the positive parity and negative  parity, respectively, the $M_{Z^\pm}$ and $\lambda_{Z^\pm}$ are the masses and pole residues, respectively.

We  insert  a complete set of intermediate hadronic states with
the same quantum numbers as the current operators   into the
correlation functions   to obtain the hadronic representation
\cite{SVZ79,Reinders85}, and isolate the ground state
tetraquark contributions,
\begin{eqnarray}
\Pi(p)&=&\frac{\lambda_{Z^+}^2}{M_{Z^+}^2-p^2} +\cdots \nonumber\\
&=&\Pi(p^2) \, ,\nonumber\\
\Pi_{\mu\nu\alpha\beta}(p)&=&\frac{\lambda_{ Z^-}^2}{M_{Z^-}^2\left(M_{Z^-}^2-p^2\right)}\left(p^2g_{\mu\alpha}g_{\nu\beta} -p^2g_{\mu\beta}g_{\nu\alpha} -g_{\mu\alpha}p_{\nu}p_{\beta}-g_{\nu\beta}p_{\mu}p_{\alpha}+g_{\mu\beta}p_{\nu}p_{\alpha}+g_{\nu\alpha}p_{\mu}p_{\beta}\right) \nonumber\\
&&+\frac{\lambda_{ Z^+}^2}{M_{Z^+}^2\left(M_{Z^+}^2-p^2\right)}\left( -g_{\mu\alpha}p_{\nu}p_{\beta}-g_{\nu\beta}p_{\mu}p_{\alpha}+g_{\mu\beta}p_{\nu}p_{\alpha}+g_{\nu\alpha}p_{\mu}p_{\beta}\right) +\cdots  \nonumber\\
&=&\widetilde{\Pi}_{-}(p^2)\left(p^2g_{\mu\alpha}g_{\nu\beta} -p^2g_{\mu\beta}g_{\nu\alpha} -g_{\mu\alpha}p_{\nu}p_{\beta}-g_{\nu\beta}p_{\mu}p_{\alpha}+g_{\mu\beta}p_{\nu}p_{\alpha}+g_{\nu\alpha}p_{\mu}p_{\beta}\right) \nonumber\\
&&+\widetilde{\Pi}_{+}(p^2)\left( -g_{\mu\alpha}p_{\nu}p_{\beta}-g_{\nu\beta}p_{\mu}p_{\alpha}+g_{\mu\beta}p_{\nu}p_{\alpha}+g_{\nu\alpha}p_{\mu}p_{\beta}\right) \, .
\end{eqnarray}

We can project out the components $\Pi_{\pm}(p^2)$ explicitly by introducing the operators $P_{Z^\pm}^{\mu\nu\alpha\beta}$,
\begin{eqnarray}
\Pi_{\pm}(p^2)&=&p^2\widetilde{\Pi}_{\pm}(p^2)=P_{Z^\pm}^{\mu\nu\alpha\beta}\Pi_{\mu\nu\alpha\beta}(p) \, ,
\end{eqnarray}
where
\begin{eqnarray}
P_{Z^-}^{\mu\nu\alpha\beta}&=&\frac{1}{6}\left( g^{\mu\alpha}-\frac{p^\mu p^\alpha}{p^2}\right)\left( g^{\nu\beta}-\frac{p^\nu p^\beta}{p^2}\right)\, , \nonumber\\
P_{Z^+}^{\mu\nu\alpha\beta}&=&\frac{1}{6}\left( g^{\mu\alpha}-\frac{p^\mu p^\alpha}{p^2}\right)\left( g^{\nu\beta}-\frac{p^\nu p^\beta}{p^2}\right)-\frac{1}{6}g^{\mu\alpha}g^{\nu\beta}\, .
\end{eqnarray}
In this article, we choose  the correlation functions $\Pi(p^2)$, $\Pi_{-}(p^2)$ and $\Pi_{+}(p^2)$ to study the scalar, vector and axialvector tetraquark states, respectively.

If  we cannot project out the components $\Pi_{\pm}(p^2)$ explicitly, we have to choose the currents $J^{\widetilde{V}}_{\mu\nu}(x)$ and $J^{\widetilde{A}}_{\mu\nu}(x)$,
\begin{eqnarray}
J^{\widetilde{V}}_{\mu\nu}(x)&=&\frac{\varepsilon^{ijk}\varepsilon^{imn}}{\sqrt{2}}\Big[u^{Tj}(x)C\gamma_5 c^k(x)  \bar{d}^m(x)\sigma^t_{\mu\nu} C \bar{c}^{Tn}(x)- u^{Tj}(x)C\sigma^t_{\mu\nu} c^k(x)  \bar{d}^m(x)\gamma_5 C \bar{c}^{Tn}(x) \Big] \, , \nonumber\\
J^{\widetilde{A}}_{\mu\nu}(x)&=&\frac{\varepsilon^{ijk}\varepsilon^{imn}}{\sqrt{2}}\Big[u^{Tj}(x)C\gamma_5 c^k(x)  \bar{d}^m(x)\sigma^v_{\mu\nu} C \bar{c}^{Tn}(x)- u^{Tj}(x)C\sigma^v_{\mu\nu} c^k(x)  \bar{d}^m(x)\gamma_5 C \bar{c}^{Tn}(x) \Big] \, , \nonumber\\
\end{eqnarray}
which couple to the $J^P=1^-$ and $1^+$ tetraquark states respectively,
\begin{eqnarray}
   \langle 0|J^{\widetilde{V}}_{\mu\nu}(0)|Z_c^-(p)\rangle &=& \frac{\widetilde{\lambda}_{Z^-}}{M_{Z^-}} \, \left(\varepsilon_{\mu}p_{\nu}-\varepsilon_{\nu}p_{\mu} \right)\, , \nonumber\\
 \langle 0|J^{\widetilde{A}}_{\mu\nu}(0)|Z_c^+(p)\rangle &=&\frac{\widetilde{\lambda}_{Z^+}}{M_{Z^+}} \left(\varepsilon_{\mu}p_{\nu}-\varepsilon_{\nu}p_{\mu} \right)\, ,
\end{eqnarray}
as  the current  operators $J^{\widetilde{V}}_{\mu\nu}(x)$ and $J^{\widetilde{A}}_{\mu\nu}(x)$ have the  properties,
\begin{eqnarray}
\widehat{P}J^{\widetilde{V}}_{\mu\nu}(x)\widehat{P}^{-1}&=&-J^{\widetilde{V}}_{\mu\nu}(\tilde{x}) \, , \nonumber\\
\widehat{P}J^{\widetilde{A}}_{\mu\nu}(x)\widehat{P}^{-1}&=&+J^{\widetilde{A}}_{\mu\nu}(\tilde{x}) \, ,
\end{eqnarray}
under parity transform $\widehat{P}$.

It is more easy to carry out the operator product expansion for the current $J_{\mu\nu}(x)$ than for the currents $J^{\widetilde{V}}_{\mu\nu}(x)$ and  $J^{\widetilde{A}}_{\mu\nu}(x)$. In this article, we prefer the current $J_{\mu\nu}(x)$, and denote the corresponding vector and axialvector tetraquark states as $S\tilde{V}$ type and $S\tilde{A}$ type, respectively.

We carry out the operator product expansion for the correlation functions up to the vacuum condensates of dimension $10$ in a consistent way, then obtain the QCD spectral densities through dispersion relation, take the
quark-hadron duality below the continuum threshold  $s_0$ and perform Borel transform  with respect to
 $P^2=-p^2$ to obtain  the  QCD sum rules:
\begin{eqnarray}\label{QCDSR}
\lambda^2_{Z}\, \exp\left(-\frac{M^2_{Z}}{T^2}\right)= \int_{4m_c^2}^{s_0} ds\, \rho(s) \, \exp\left(-\frac{s}{T^2}\right) \, ,
\end{eqnarray}
the $\rho(s)$ are the QCD spectral densities. The explicit expressions of the QCD spectral densities are available upon request by contacting me with E-mail. For the technical details, one can consult Refs.\cite{WangHuangtao-2014-PRD,Wang-tetra-formula}. In carrying out the operator product expansion for the correlation functions
$\Pi(p)$, we encounter    the terms $(p\cdot t)^2$, and set $(p\cdot t)^2=p^2$, just like in the QCD sum rules for the baryon states  separating the contributions of the positive parity and negative parity, where we take the four vector $p^\mu=(p_0,\vec{0})$ \cite{p0-Baryon}.

We derive Eq.\eqref{QCDSR} with respect to  $\tau=\frac{1}{T^2}$,  and obtain the QCD sum rules for
 the masses of the  scalar, vector and axialvector   tetraquark states $Z_c$ through a ratio,
 \begin{eqnarray}\label{QCDSR-mass}
 M^2_{Z}&=& -\frac{\int_{4m_c^2}^{s_0} ds\frac{d}{d \tau}\rho(s)\exp\left(-\tau s \right)}{\int_{4m_c^2}^{s_0} ds \rho(s)\exp\left(-\tau s\right)}\, .
\end{eqnarray}

\section{Numerical results and discussions}
We choose  the standard values of the vacuum condensates $\langle
\bar{q}q \rangle=-(0.24\pm 0.01\, \rm{GeV})^3$,   $\langle
\bar{q}g_s\sigma G q \rangle=m_0^2\langle \bar{q}q \rangle$,
$m_0^2=(0.8 \pm 0.1)\,\rm{GeV}^2$,  $\langle \frac{\alpha_s
GG}{\pi}\rangle=(0.33\,\rm{GeV})^4 $    at the energy scale  $\mu=1\, \rm{GeV}$
\cite{SVZ79,Reinders85,Colangelo-Review}, and choose the $\overline{MS}$ mass $m_{c}(m_c)=(1.275\pm0.025)\,\rm{GeV}$ from the Particle Data Group \cite{PDG}, and set $m_u=m_d=0$.
Moreover, we take into account the energy-scale dependence of  the input parameters at  the QCD side,
\begin{eqnarray}
\langle\bar{q}q \rangle(\mu)&=&\langle\bar{q}q \rangle({\rm 1GeV})\left[\frac{\alpha_{s}({\rm 1GeV})}{\alpha_{s}(\mu)}\right]^{\frac{12}{25}}\, , \nonumber\\
 \langle\bar{q}g_s \sigma Gq \rangle(\mu)&=&\langle\bar{q}g_s \sigma Gq \rangle({\rm 1GeV})\left[\frac{\alpha_{s}({\rm 1GeV})}{\alpha_{s}(\mu)}\right]^{\frac{2}{25}}\, , \nonumber\\ m_c(\mu)&=&m_c(m_c)\left[\frac{\alpha_{s}(\mu)}{\alpha_{s}(m_c)}\right]^{\frac{12}{25}} \, ,\nonumber\\
\alpha_s(\mu)&=&\frac{1}{b_0t}\left[1-\frac{b_1}{b_0^2}\frac{\log t}{t} +\frac{b_1^2(\log^2{t}-\log{t}-1)+b_0b_2}{b_0^4t^2}\right]\, ,
\end{eqnarray}
   where $t=\log \frac{\mu^2}{\Lambda^2}$, $b_0=\frac{33-2n_f}{12\pi}$, $b_1=\frac{153-19n_f}{24\pi^2}$, $b_2=\frac{2857-\frac{5033}{9}n_f+\frac{325}{27}n_f^2}{128\pi^3}$,  $\Lambda=210\,\rm{MeV}$, $292\,\rm{MeV}$  and  $332\,\rm{MeV}$ for the flavors  $n_f=5$, $4$ and $3$, respectively  \cite{PDG,Narison-mix}, and evolve all the input parameters to the ideal energy scales   $\mu$ to extract the tetraquark masses. In the present work, we choose the flavor  $n_f=4$.

 Now we search for the ideal  Borel parameters $T^2$ and continuum threshold parameters $s_0$  to satisfy   the  four criteria:\\
$\bf 1.$ Pole dominance at the hadron side;\\
$\bf 2.$ Convergence of the operator product expansion;\\
$\bf 3.$ Appearance of the Borel platforms;\\
$\bf 4.$ Satisfying the energy scale formula,\\
 via  try and error. The resulting Borel parameters, continuum threshold parameters, energy scales of the QCD spectral densities and pole contributions are shown explicitly in Table \ref{BorelP}.
From the Table,  we can see that the pole contributions are about $(40-60)\%$, the pole dominance condition at the hadron side is well satisfied. In calculations, we observe that the contributions of the vacuum condensates of dimension $10$ are $\ll 1\%$, the operator product expansion is well convergent.

\begin{figure}
\centering
\includegraphics[totalheight=6cm,width=7cm]{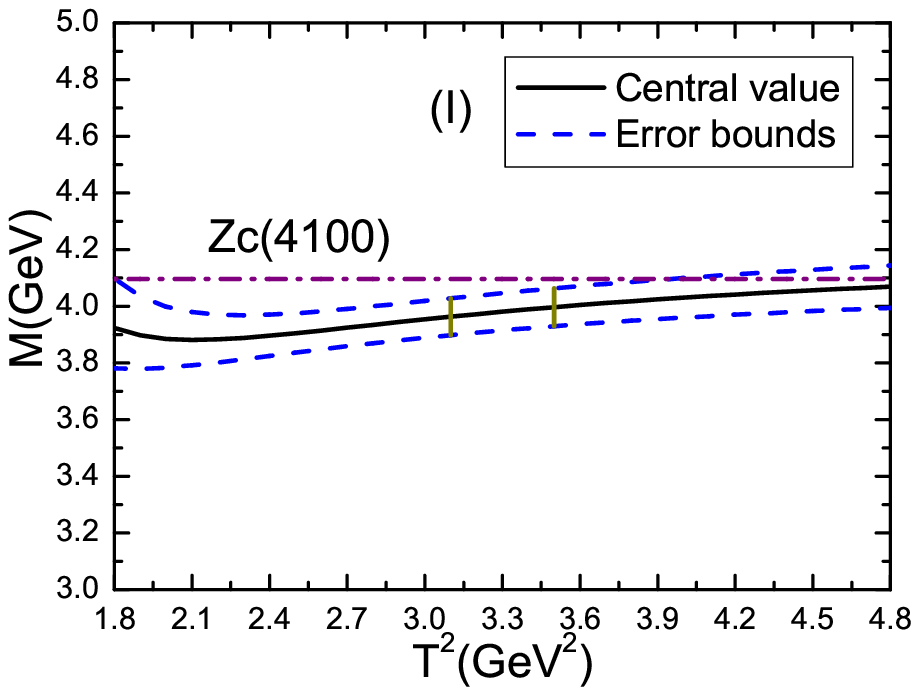}
\includegraphics[totalheight=6cm,width=7cm]{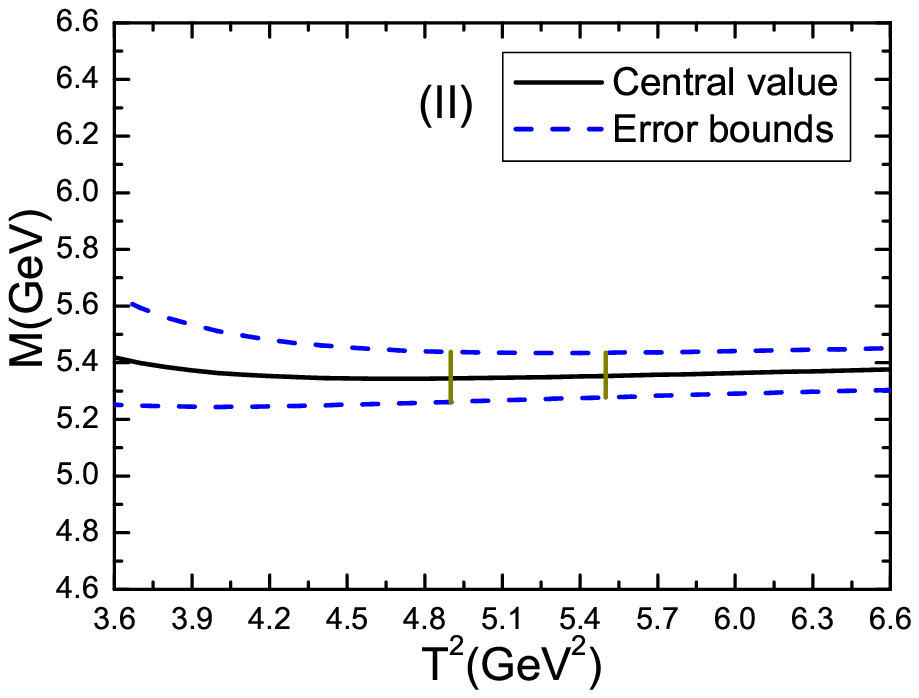}
\includegraphics[totalheight=6cm,width=7cm]{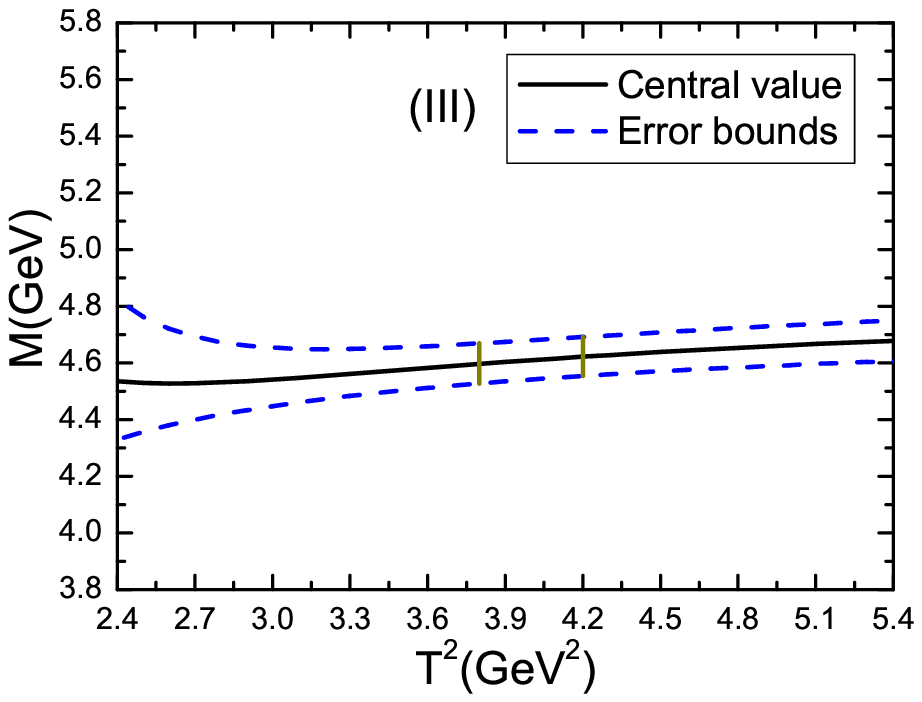}
\includegraphics[totalheight=6cm,width=7cm]{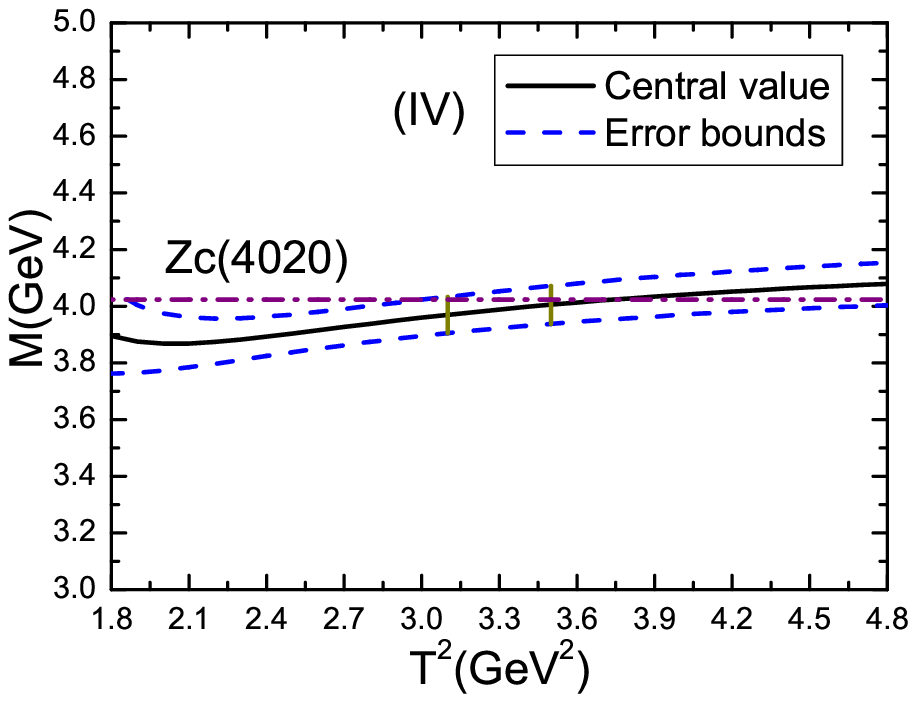}
  \caption{ The masses  with variations of the  Borel parameters $T^2$ for  the tetraquark states, the (I), (II), (III) and (IV) denote the $[uc]_{\tilde{A}}[\bar{d}\bar{c}]_{\tilde{A}}$, $[uc]_{\tilde{V}}[\bar{d}\bar{c}]_{\tilde{V}}$, $[uc]_S[\bar{d}\bar{c}]_{\tilde{V}}-[uc]_{\tilde{V}}[\bar{d}\bar{c}]_S$
and $[uc]_S[\bar{d}\bar{c}]_{\tilde{A}}-[uc]_{\tilde{A}}[\bar{d}\bar{c}]_S$  tetraquark states, respectively, the regions between the two
     perpendicular lines are the Borel windows. }\label{mass-fig}
\end{figure}

\begin{figure}
\centering
\includegraphics[totalheight=6cm,width=7cm]{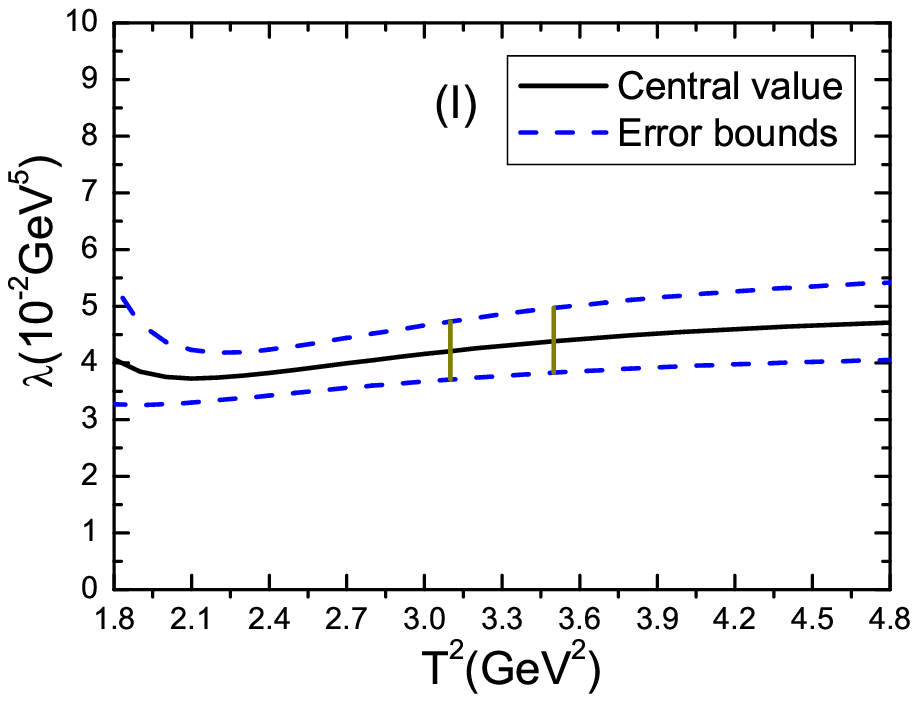}
\includegraphics[totalheight=6cm,width=7cm]{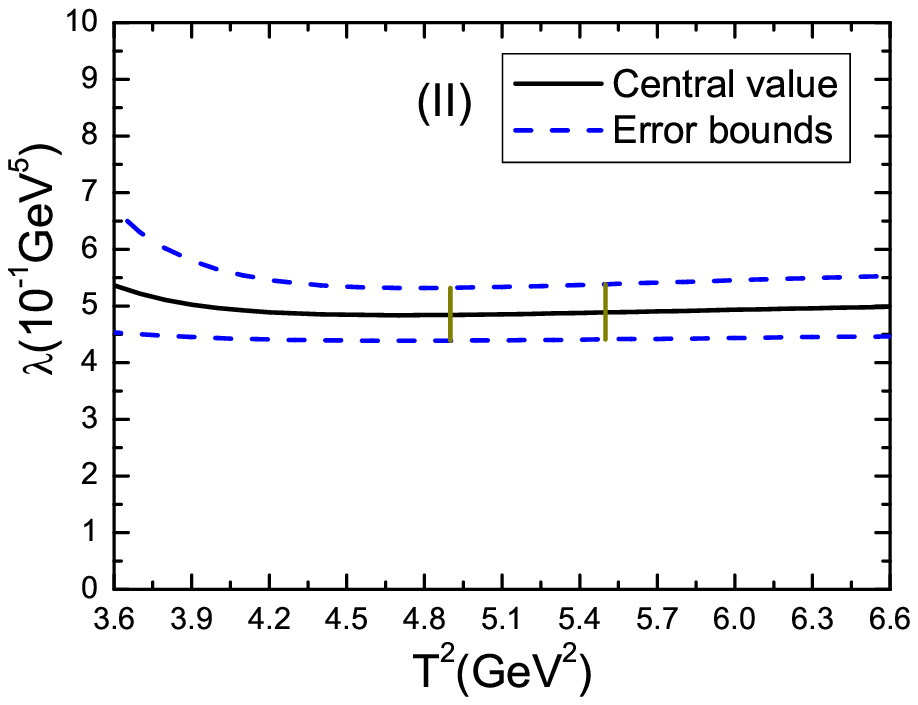}
\includegraphics[totalheight=6cm,width=7cm]{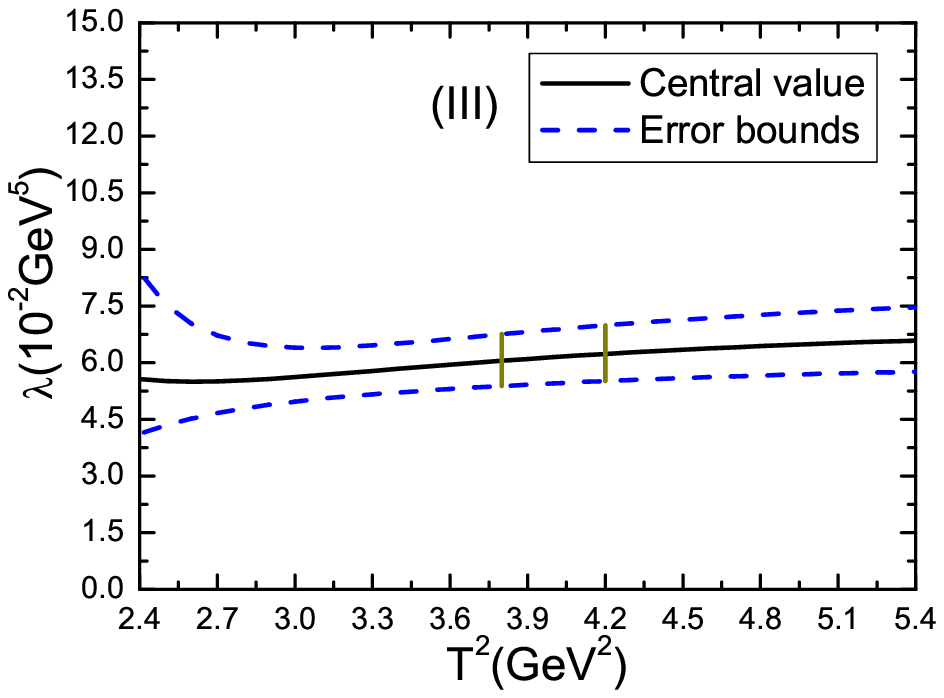}
\includegraphics[totalheight=6cm,width=7cm]{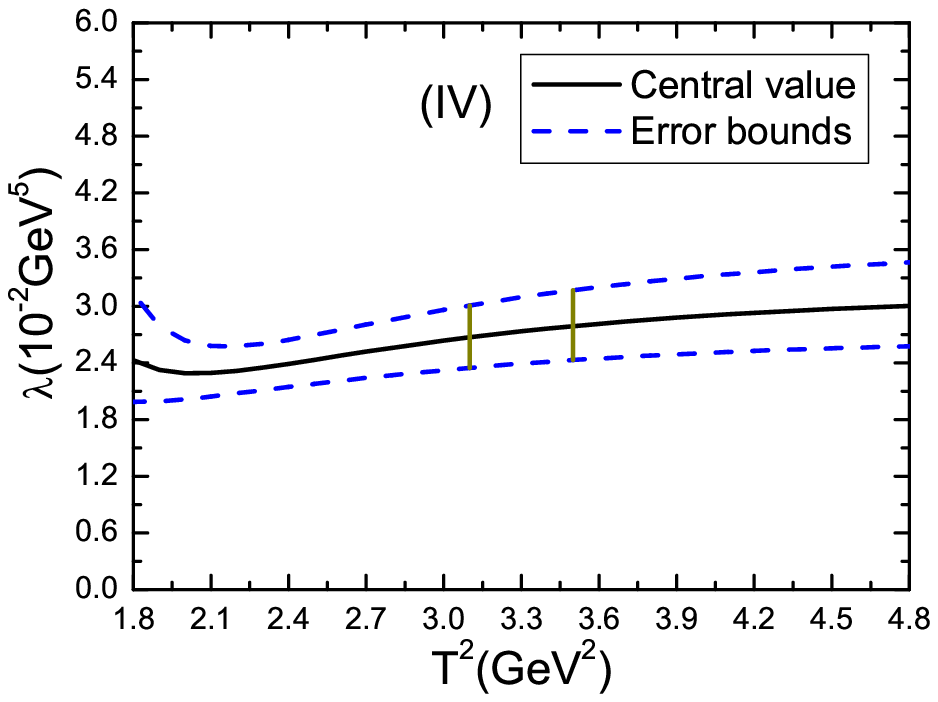}
  \caption{ The pole residues  with variations of the  Borel parameters $T^2$ for  the tetraquark states, the (I), (II), (III) and (IV) denote the $[uc]_{\tilde{A}}[\bar{d}\bar{c}]_{\tilde{A}}$, $[uc]_{\tilde{V}}[\bar{d}\bar{c}]_{\tilde{V}}$, $[uc]_S[\bar{d}\bar{c}]_{\tilde{V}}-[uc]_{\tilde{V}}[\bar{d}\bar{c}]_S$
and $[uc]_S[\bar{d}\bar{c}]_{\tilde{A}}-[uc]_{\tilde{A}}[\bar{d}\bar{c}]_S$  tetraquark states, respectively, the regions between the two
     perpendicular lines are the Borel windows. } \label{residue-fig}
\end{figure}

We take into account the uncertainties of the input parameters and obtain the masses and pole residues of the tetraquark states, which are shown explicitly in Table \ref{mass-Table} and in Figs.\ref{mass-fig}--\ref{residue-fig}. From  Tables \ref{BorelP}--\ref{mass-Table}, we can see that the energy scale formula $\mu=\sqrt{M^2_{X/Y/Z}-(2{\mathbb{M}}_c)^2}$ is well satisfied, where we take the updated value of the effective $c$-quark mass  ${\mathbb{M}}_c=1.82\,\rm{GeV}$ \cite{WangEPJC-1601}. In  Figs.\ref{mass-fig}--\ref{residue-fig}, we plot the masses and pole residues of the tetraquark states with variations of the Borel parameters at much larger ranges than the Borel widows, the regions between the two
    perpendicular lines are the Borel windows. In the Borel windows, the uncertainties induced by the Borel parameters are $\ll1\%$ for the masses and $\leq2\%$ for the pole residues, there appear Borel platforms. Now the four criteria are all satisfied, we expect to make reliable predictions.

In Ref.\cite{WangZhang-tt},  we study the   tensor-tensor type scalar hidden-charm tetraquark states with currents
\begin{eqnarray}
\eta(x)&=&\varepsilon^{ijk}\varepsilon^{imn}q^{Tj}(x)C\sigma_{\mu\nu} c^k(x)  \bar{q}^m(x)\sigma^{\mu\nu} C \bar{c}^{Tn}(x) \, ,
\end{eqnarray}
 via the QCD sum rules by taking into account  both the ground state contributions and the first radial excited state contributions, where $\sigma_{\mu\nu}=\frac{i}{2}[\gamma_\mu,\gamma_\nu]=\sigma^t_{\mu\nu}+\sigma^v_{\mu\nu}$, and   obtain masses
$M_{TT,{\rm 1S}}=3.82\pm0.16 \,\rm{GeV}$ and $M_{TT,{\rm 2S}}=4.38\pm0.09 \,\rm{GeV}$. We can rewrite the current $\eta(x)$ as a linear  superposition $\eta(x)=J_{\tilde{V}\tilde{V}}(x)+2J_{\tilde{A}\tilde{A}}(x)$, the tensor-tensor type scalar hidden-charm tetraquark states have both
 the $\tilde{V}\tilde{V}$ and $\tilde{A}\tilde{A}$ components. Compared to the $\tilde{A}\tilde{A}$ ($\tilde{V}\tilde{V}$) type tetraquark mass $M_{\tilde{A}\tilde{A}}=3.98\pm0.08\,\rm{GeV}$ ($M_{\tilde{V}\tilde{V}}=5.35\pm0.09\,\rm{GeV}$), the ground state mass
 $M_{TT,{\rm 1S}}=3.82\pm0.16 \,\rm{GeV}$ is (much) lower.  In the QCD sum rules for the tensor-tensor type scalar tetraquark states, the terms $m_c\langle \bar{q}q\rangle$ and $m_c\langle \bar{q}g_s\sigma Gq\rangle$ disappear due to the special superposition $\eta(x)=J_{\tilde{V}\tilde{V}}(x)+2J_{\tilde{A}\tilde{A}}(x)$, the contamination from  the $\tilde{V}\tilde{V}$ component  is large.

 In Fig.\ref{mass-fig}, we also present the experimental values of the masses of the $Z_c(4100)$ and $Z_c(4020)$ \cite{LHCb-Z4100,PDG,BES1308,BES1309}. From the figure, we can see that the predicted mass for the $[uc]_S[\bar{d}\bar{c}]_{\tilde{A}}-[uc]_{\tilde{A}}[\bar{d}\bar{c}]_S$ type axialvector tetraquark state is in excellent agreement with  the experimental data in the Borel  window,
and favors assigning the $Z_c(4020)$ to be the $[uc]_S[\bar{d}\bar{c}]_{\tilde{A}}-[uc]_{\tilde{A}}[\bar{d}\bar{c}]_S$ type axialvector tetraquark state,  while the $Z_c(4100)$ lies above the $\tilde{A}\tilde{A}$-type scalar tetraquark state, and much below the $\tilde{V}\tilde{V}$-type scalar tetraquark state and $S\tilde{V}$-type vector tetraquark state in the Borel windows, the present QCD sum rules do not favor assigning the $Z_c(4100)$ to be the scalar or vector tetraquark state.
 The masses of the scalar hidden-charm tetraquark states have the hierarchy \cite{Wang-3915-CgmCgm,Wang-3915-C5C5,Wang-3860},
 \begin{eqnarray}\label{mass-relation}
 M_{SS}\leq M_{AA} \leq M_{\tilde{A}\tilde{A}}< M_{Z_c(4100)} \ll M_{VV}\ll M_{\tilde{V}\tilde{V}}\leq M_{PP} \, ,
\end{eqnarray}
the QCD sum rules disfavor assigning the $Z_c(4100)$ to be a scalar tetraquark state.

The masses extracted from the QCD sum rules depend on the Borel windows, different Borel windows lead to different predicted masses.
From Fig.\ref{mass-fig}(I), we can see that if we choose larger Borel parameter for the $\tilde{A}\tilde{A}$-type scalar tetraquark state, for example, choose $T^2>4.2\,\rm{GeV}^2$ rather than choose $T^2=(3.1-3.5)\,\rm{GeV}^2$, we can obtain a mass about $4.1\,\rm{GeV}$, which is compatible with the experimental value of the mass of the $Z_c(4100)$.
If we choose the Borel window $T^2=(4.9-5.3)\,\rm{GeV}^2$, the predicted mass $M_{\tilde{A}\tilde{A}}=4.09\pm0.08\,\rm{GeV}$, which is in excellent agreement with the experimental data $M_{Z_c}=4096 \pm 20^{+18}_{-22}\,\rm{MeV}$ \cite{LHCb-Z4100}. However, the pole contribution is about $(14-24)\%$, the prediction is not robust.

\begin{figure}
\centering
\includegraphics[totalheight=7cm,width=9cm]{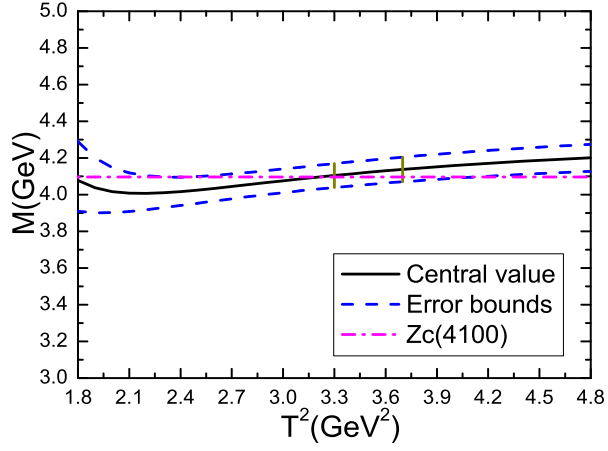}
  \caption{ The mass  with variation of the  Borel parameter $T^2$ for  the $[uc]_{\tilde{A}}[\bar{d}\bar{c}]_{\tilde{A}}$ type scalar tetraquark state at the energy scale $\mu=1.4\,\rm{GeV}$, the region between the two  perpendicular line is the Borel window. }\label{mass-14-fig}
\end{figure}

The  mass $M_{\tilde{A}\tilde{A}}$ extracted from the QCD sum rules decreases  monotonously  with increase of the energy scales of the QCD spectral density. If we choose $\mu=1.4\,\rm{GeV}$, $T^2=(3.3-3.7)\,\rm{GeV}^2$, $\sqrt{s_0}=4.70\pm0.10\,\rm{GeV}$, the pole contribution is $(41-61)\%$ and the operator product expansion is well convergent, we obtain the tetraquark  mass $M_{\tilde{A}\tilde{A}}=4.12\pm0.08\,\rm{GeV}$, which is in excellent agreement with the experimental data $M_{Z_c}=4096 \pm 20^{+18}_{-22}\,\rm{MeV}$ \cite{LHCb-Z4100}, see Fig.\ref{mass-14-fig}. In Fig.\ref{mass-14-fig}, we plot the  mass $M_{\tilde{A}\tilde{A}}$ extracted at the energy scale $\mu=1.4\,\rm{GeV}$ with variation of the Borel parameter, again the region between the two
     perpendicular lines is the Borel windows. However, the energy scale formula $\mu=\sqrt{M^2_{X/Y/Z}-(2{\mathbb{M}}_c)^2}$ is not satisfied.
In the QCD sum rules for the hidden-charm (or hidden-bottom) tetraquark states and molecular states, the integrals
 \begin{eqnarray}
 \int_{4m_Q^2(\mu)}^{s_0} ds \rho_{QCD}(s,\mu)\exp\left(-\frac{s}{T^2} \right)\, ,
 \end{eqnarray}
are sensitive to the energy scales $\mu$.  We suggest an energy scale formula    $\mu=\sqrt{M^2_{X/Y/Z}-(2{\mathbb{M}}_Q)^2}$ with the effective $Q$-quark mass ${\mathbb{M}}_Q$ to determine the ideal energy scales of the QCD spectral densities in a consistent way \cite{Wang-tetra-formula}. The energy scale formula works well for the tetraquark states \cite{WangHuangtao-2014-PRD, Wang-3915-CgmCgm, Wang-3915-C5C5, Wang-3860,WangEPJC-1601,Wang-tetra-formula,Wang-Z4600-etal,Wang-tetra-IJMPA},  tetraquark molecular states \cite{WangHuang-molecule}, and even for the hidden-charm pentaquark states \cite{WangPentaQuark}. For example, in Refs.\cite{WangHuangtao-2014-PRD,WangEPJC-1601,Wang-Z4600-etal} and the present work, we observe that there exist one axialvector tetraquark candidate $[uc]_S[\bar{d}\bar{c}]_A-[uc]_A[\bar{d}\bar{c}]_S$ for the  $Z_c(3900)$,  three axialvector tetraquark candidates $[uc]_A[\bar{d}\bar{c}]_A $, $[uc]_{\tilde{A}}[\bar{d}\bar{c}]_A-[uc]_A[\bar{d}\bar{c}]_{\tilde{A}}$  and $[uc]_S[\bar{d}\bar{c}]_{\tilde{A}}-[uc]_{\tilde{A}}[\bar{d}\bar{c}]_S$ for the $Z_c(4020)$, which is consistent with  the almost degenerate scalar and axialvector heavy diquark  masses from the QCD sum rules \cite{WangDiquark}. Furthermore, the $Z_c(4430)$ can be assigned to be the first radial excited state of the $Z_c(3900)$. If the $Z_c(4100)$ is a diquark-antidiquark type scalar tetraquark state, the energy scale formula should be satisfied, as the $Z_c(4100)$ has no reason to be an odd or special tetraquark state.  The masses of the scalar tetraquark states $SS$, $AA$ and $\tilde{A}\tilde{A}$ are estimated to be $3.9-4.0\,\rm{GeV}$, if the spin-spin interactions are neglected, the QCD sum rules support this estimation.

In Ref.\cite{Azizi-Zc4100}, Sundu,   Agaev and  Azizi choose the $SS$ type scalar current to study the mass and width of the $Z_c(4100)$, and obtain the mass $M_{SS}=4.08\pm0.15\,\rm{GeV}$ with the pole contribution $(19-54)\%$. In Ref.\cite{Wang-3860}, we tentatively assign the $X^*(3860)$ to be the $[qc]_S[\bar{q}\bar{c}]_S$-type  scalar
tetraquark state, study its mass and width with the QCD sum rules, and obtain the mass $M_{SS} = 3.86 \pm 0.09\,\rm{GeV}$ with the pole contribution  $(46-70)\%$, which is much larger  than the pole contribution $(19-54)\%$ in Ref.\cite{Azizi-Zc4100}. In Ref.\cite{Wang-3860}, just like in the present work, we use the energy scale formula $\mu=\sqrt{M^2_{X/Y/Z}-(2{\mathbb{M}}_c)^2}$ to enhance the pole contribution. In the QCD sum rules, we prefer larger pole contributions to obtain more robust predictions.

\begin{figure}
\centering
\includegraphics[totalheight=7cm,width=9cm]{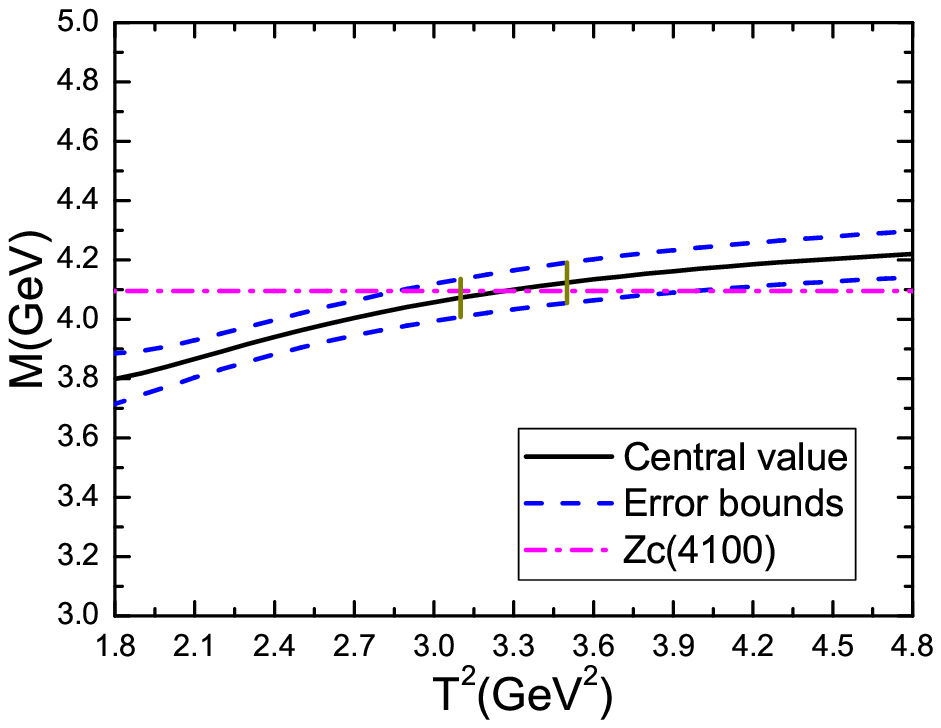}
  \caption{ The mass  with variation of the  Borel parameter $T^2$ for  the $[uc]_{\tilde{A}}[\bar{d}\bar{c}]_{\tilde{A}}\oplus[uc]_{\tilde{V}}[\bar{d}\bar{c}]_{\tilde{V}}$ type scalar  tetraquark state, the region between the two  perpendicular line is the Borel window. }\label{mass-AA-VV-fig}
\end{figure}

Now we assume that  the $Z_c(4100)$ is a $\tilde{A}\tilde{A}\oplus\tilde{V}\tilde{V}$ type mixing scalar tetraquark state, and  study it with the current $J(x)$,
\begin{eqnarray}
J(x)&=&2J_{\tilde{A}\tilde{A}}(x)\cos\theta+J_{\tilde{V}\tilde{V}}(x)\sin\theta \, ,
\end{eqnarray}
where we introduce the factor $2$ due to the relation $\sigma_{\mu\nu}=\sigma^t_{\mu\nu}+\sigma^v_{\mu\nu}$. If we choose the mixing angle $\theta=26.4^\circ$, the energy scale $\mu=1.9\,\rm{GeV}$, the Borel parameter $T^2=(3.1-3.5)\,\rm{GeV}^2$, the continuum threshold parameter $\sqrt{s_0}=4.70\pm0.10\,\rm{GeV}$, the pole contribution is $(42-62)\%$,  the operator product expansion is well convergent, the energy scale formula is also satisfied. We obtain the tetraquark  mass $M_{\tilde{A}\tilde{A}\oplus\tilde{V}\tilde{V}}=4.10\pm0.09\,\rm{GeV}$, which is in excellent agreement with the experimental data $M_{Z_c}=4096 \pm 20^{+18}_{-22}\,\rm{MeV}$ \cite{LHCb-Z4100}, see Fig.\ref{mass-AA-VV-fig}. If the $Z_c(4100)$ is a diquark-antidiquark type  tetraquark state, it may be a $\tilde{A}\tilde{A}\oplus\tilde{V}\tilde{V}$ type mixing scalar tetraquark state.

We can also introduce more parameters and write down the most general scalar current $J(x)$,
\begin{eqnarray}
J(x)&=&2J_{\tilde{A}\tilde{A}}(x)\,t_1+J_{\tilde{V}\tilde{V}}(x)\,t_2+  J_{SS}(x)\,t_3+ J_{PP}(x)\,t_4+J_{AA}(x)\,t_5+J_{VV}(x)\,t_6\, ,
\end{eqnarray}
where
\begin{eqnarray}
 J_{SS}(x)&=&\varepsilon^{ijk}\varepsilon^{imn}u^T_j(x)C\gamma_5 c_k(x)\, \bar{d}_m(x)\gamma_5 C \bar{c}^T_n(x) \, ,  \nonumber \\
  J_{PP}(x)&=&\varepsilon^{ijk}\varepsilon^{imn}u^T_j(x)C c_k(x)\, \bar{d}_m(x) C \bar{c}^T_n(x) \, ,  \nonumber \\
 J_{AA}(x)&=&\varepsilon^{ijk}\varepsilon^{imn}u^T_j(x)C\gamma_\mu c_k(x)\, \bar{d}_m(x)\gamma^\mu C \bar{c}^T_n(x) \, ,  \nonumber \\
  J_{VV}(x)&=&\varepsilon^{ijk}\varepsilon^{imn}u^T_j(x)C\gamma_\mu\gamma_5 c_k(x) \,\bar{d}_m(x)\gamma_5\gamma^\mu C \bar{c}^T_n(x) \, ,
\end{eqnarray}
the $t_i$ with $i=1$, $2$, $3$, $\cdots$ are arbitrary  parameters. We can obtain any value  between the largest mass $M_{PP}$ and the smallest mass $M_{SS}$ by fine tuning the parameters $t_i$, see Eq.\eqref{mass-relation}. In fact, we cannot assign a tetraquark state  unambiguously with the mass alone, we have to study the decays exclusively to obtain the partial decay widths and confront the predictions to experimental data in the future. The cumbersome calculations  may be our next work.
In the present work, we can obtain the conclusion tentatively that we cannot reproduce the experimental value of the mass of $Z_c(4100)$ without introducing mixing effects.

In this article, we take the zero width approximation. In fact, we can take into account the finite width effect with the  simple replacement of the hadronic spectral density,
\begin{eqnarray}
\lambda^2_{Z_c}\delta \left(s-M^2_{Z_c} \right) &\to& \lambda^2_{Z_c}\frac{1}{\pi}\frac{M_{Z_c}\Gamma_{Z_c}(s)}{(s-M_{Z_c}^2)^2+M_{Z_c}^2\Gamma_{Z_c}^2(s)}\, ,
\end{eqnarray}
where
\begin{eqnarray}
\Gamma_{Z_c}(s)&=&\Gamma_{Z_c} \frac{M_{Z_c}}{\sqrt{s}}\sqrt{\frac{\left[s-(M_{\eta_c}+M_{\pi})^2\right]\left[s-(M_{\eta_c}-M_{\pi})^2\right]}{\left[M^2_{Z_c}-(M_{\eta_c}+M_{\pi})^2\right]\left[M^2_{Z_c}-(M_{\eta_c}-M_{\pi})^2\right]}} \, .
\end{eqnarray}
Then the hadron sides of  the QCD sum rules in Eqs.\eqref{QCDSR}-\eqref{QCDSR-mass} undergo the  changes,
\begin{eqnarray}
\lambda^2_{Z_c}\exp \left(-\frac{M^2_{Z_c}}{T^2} \right) &\to& \lambda^2_{Z_c}\int_{(M_{\eta_c}+M_{\pi})^2}^{s_0}ds\frac{1}{\pi}\frac{M_{Z_c}\Gamma_{Z_c}(s)}{(s-M_{Z_c}^2)^2+M_{Z_c}^2\Gamma_{Z_c}^2(s)}\exp \left(-\frac{s}{T^2} \right)\, , \nonumber\\
&=&0.96\,\lambda^2_{Z_c}\exp \left(-\frac{M^2_{Z_c}}{T^2} \right)\, , \\
\lambda^2_{Z_c}M^2_{Z_c}\exp \left(-\frac{M^2_{Z_c}}{T^2} \right) &\to& \lambda^2_{Z_c}\int_{(M_{\eta_c}+M_{\pi})^2}^{s_0}ds\,s\,\frac{1}{\pi}\frac{M_{Z_c}\Gamma_{Z_c}(s)}{(s-M_{Z_c}^2)^2+M_{Z_c}^2\Gamma_{Z_c}^2(s)}\exp \left(-\frac{s}{T^2} \right)\, , \nonumber\\
&=&0.94\,\lambda^2_{Z_c}M^2_{Z_c}\exp \left(-\frac{M^2_{Z_c}}{T^2} \right)\, ,
\end{eqnarray}
where we have used the central values of the input parameters for the $[uc]_{\tilde{A}}[\bar{d}\bar{c}]_{\tilde{A}}\oplus[uc]_{\tilde{V}}[\bar{d}\bar{c}]_{\tilde{V}}$ type scalar tetraquark state shown in Table \ref{BorelP} and the physical values of the mass and width of the $Z_c(4100)$. We can absorb the numerical factors  $0.96$ and $0.94$ into the pole residue $\lambda_{Z_c}$ safely with the simple replacement $\lambda_{Z_c} \to (0.97-0.98)\lambda_{Z_c}$, the zero width approximation works well.

The predicted mass of the $[uc]_S[\bar{d}\bar{c}]_{\tilde{V}}-[uc]_{\tilde{V}}[\bar{d}\bar{c}]_S$ type vector  tetraquark state $M_{S\tilde{V}}=4.61\pm0.08\,\rm{GeV}$ is much larger than the mass of the $Z_c(4100)$, see Table \ref{mass-Table}. Furthermore,  the mass of the $Z_c(4100)$ is even smaller than the lowest vector hidden-charm  tetraquark mass  from the QCD sum rules \cite{Wang-Vector-Zc4100,Wang-Vector-tetraquark},
\begin{eqnarray}
M_{Z_c(4100)}<M_{Y(4260/4220)}=4.24\pm0.10\,{\rm{GeV}}\ll M_{S\tilde{V}}=4.61\pm0.08\,\rm{GeV} \, ,
\end{eqnarray}
the QCD sum rules also disfavor assigning the $Z_c(4100)$ to be a vector  tetraquark state.

If the $Z_c(4100)$ is a $SS$, $AA$ or $\tilde{A}\tilde{A}$ type scalar hidden-charm tetraquark state without mixing effects, it should have a mass about $3.9\,\rm{GeV}$ or $4.0\,\rm{GeV}$ rather than $4.1\,\rm{GeV}$; on the other hand, if the $Z_c(4100)$ is a vector hidden-charm tetraquark state, it should have a mass about $4.2\,\rm{GeV}$ rather than $4.1\,\rm{GeV}$. However, a  $\tilde{A}\tilde{A}\oplus\tilde{V}\tilde{V}$  type mixing scalar tetraquark state can have a mass about $4.1\,\rm{GeV}$ and reproduce the experimental value of the mass of the $Z_c(4100)$ \cite{LHCb-Z4100}.

The $Z_c^-(4100)$ is observed in the $\eta_c \pi^-$ mass spectrum,
the spin-parity assignments $J^P =0^+$ and $1^-$ are both consistent with the experimental data. We can search for its  charge zero partner $Z_c^0(4100)$ in the $\eta_c \pi^0$ mass spectrum, if the $\eta_c \pi^0$ is in relative S-wave, the $Z_c^0(4100)$ has the $J^{PC}=0^{++}$, on the other hand, if the $\eta_c \pi^0$ is in relative P-wave, the $Z_c^0(4100)$ has the $J^{PC}=1^{-+}$. The quantum numbers $J^{PC}=1^{-+}$ is exotic, we can also search for the $Z_c^0(4100)$ in the $J/\psi \rho^0$ mass spectrum,  which maybe shed light on the nature of the $Z_c(4100)$. In Ref.\cite{Wang-tetra-formula}, we observe that the diquark-antidiquark type vector $cu\bar{c}\bar{d}$ (or $cd\bar{c}\bar{u}$) tetraquark state with $J^{PC}=1^{-+}$ has smaller mass than the corresponding vector tetraquark state with $J^{PC}=1^{--}$, about $0.09\,\rm{GeV}$.
In the present case, we can choose the current $\eta_{\mu\nu}(x)$,
\begin{eqnarray}
\eta_{\mu\nu}(x)&=&\frac{\varepsilon^{ijk}\varepsilon^{imn}}{\sqrt{2}}\Big[u^{Tj}(x)C\gamma_5 c^k(x)  \bar{d}^m(x)\sigma_{\mu\nu} C \bar{c}^{Tn}(x)+ u^{Tj}(x)C\sigma_{\mu\nu} c^k(x)  \bar{d}^m(x)\gamma_5 C \bar{c}^{Tn}(x) \Big]\, ,\nonumber\\
\end{eqnarray}
which couples potentially to the vector tetraquark state with $J^{PC}=1^{-+}$, the mass of the  vector tetraquark state with $J^{PC}=1^{-+}$ is estimated to be  $4.52\pm 0.08\,\rm{GeV}$, which is much larger than the experimental value of the mass of the $Z_c^-(4100)$, $4096\pm 20  {}^{+18}_{-22}\,\rm{ MeV}$.

\begin{table}
\begin{center}
\begin{tabular}{|c|c|c|c|c|c|c|c|c|}\hline\hline
 $Z_c$                                                & $T^2 (\rm{GeV}^2)$ & $\sqrt{s_0}(\rm GeV) $      & $\mu(\rm{GeV})$   & pole                 \\ \hline

$[uc]_{\tilde{A}}[\bar{d}\bar{c}]_{\tilde{A}}$        & $3.1-3.5$          & $4.55\pm0.10$               & $1.6$             & $(42-62)\%$         \\ \hline

$[uc]_{\tilde{V}}[\bar{d}\bar{c}]_{\tilde{V}}$        & $4.9-5.5$          & $5.90\pm0.10$               & $3.9$             & $(43-61)\%$         \\ \hline

$[uc]_S[\bar{d}\bar{c}]_{\tilde{V}}-[uc]_{\tilde{V}}[\bar{d}\bar{c}]_S$  & $3.8-4.2$   & $5.18\pm0.10$   & $2.8$             & $(43-61)\%$         \\ \hline

$[uc]_S[\bar{d}\bar{c}]_{\tilde{A}}-[uc]_{\tilde{A}}[\bar{d}\bar{c}]_S$  & $3.1-3.5$   & $4.56\pm0.10$   & $1.6$             & $(42-62)\%$         \\ \hline

$[uc]_{\tilde{A}}[\bar{d}\bar{c}]_{\tilde{A}}^*$     & $3.3-3.7$          & $4.70\pm0.10$                & $1.4$             & $(41-61)\%$         \\ \hline

$[uc]_{\tilde{A}}[\bar{d}\bar{c}]_{\tilde{A}}\oplus[uc]_{\tilde{V}}[\bar{d}\bar{c}]_{\tilde{V}}$   & $3.1-3.5$    & $4.70\pm0.10$  & $1.9$    & $(42-62)\%$  \\ \hline \hline
\end{tabular}
\end{center}
\caption{ The Borel parameters, continuum threshold parameters, energy scales of the QCD spectral densities and pole contributions  of the ground state tetraquark states, where the superscript ${}^*$ denotes the
 energy scale formula is not satisfied. }\label{BorelP}
\end{table}

\begin{table}
\begin{center}
\begin{tabular}{|c|c|c|c|c|c|c|c|c|}\hline\hline
 $Z_c$                                                                    & $M_Z (\rm{GeV})$   & $\lambda_Z (\rm{GeV}^5) $             \\ \hline

$[uc]_{\tilde{A}}[\bar{d}\bar{c}]_{\tilde{A}}$                            & $3.98\pm0.08$      & $(4.30\pm0.63)\times 10^{-2}$           \\ \hline

$[uc]_{\tilde{V}}[\bar{d}\bar{c}]_{\tilde{V}}$                            & $5.35\pm0.09$      & $(4.86\pm0.50)\times 10^{-1}$           \\ \hline

$[uc]_S[\bar{d}\bar{c}]_{\tilde{V}}-[uc]_{\tilde{V}}[\bar{d}\bar{c}]_S$   & $4.61\pm0.08$      & $(6.15\pm0.80)\times 10^{-2}$        \\ \hline

$[uc]_S[\bar{d}\bar{c}]_{\tilde{A}}-[uc]_{\tilde{A}}[\bar{d}\bar{c}]_S$   & $3.99\pm0.09$      & $(2.73\pm0.41)\times 10^{-2}$           \\ \hline

$[uc]_{\tilde{A}}[\bar{d}\bar{c}]_{\tilde{A}}^*$                          & $4.12\pm0.08$      & $(4.84\pm0.71)\times 10^{-2}$           \\ \hline

$[uc]_{\tilde{A}}[\bar{d}\bar{c}]_{\tilde{A}}\oplus[uc]_{\tilde{V}}[\bar{d}\bar{c}]_{\tilde{V}}$    & $4.10\pm0.09$    & $(1.43\pm0.24)\times 10^{-1}$   \\ \hline
\hline
\end{tabular}
\end{center}
\caption{ The masses and pole residues of the ground state tetraquark states, where the superscript ${}^*$ denotes the
 energy scale formula is not satisfied. }\label{mass-Table}
\end{table}

\section{Conclusion}
In this article,  we separate the vector and axialvector components of the tensor diquark operators explicitly, construct the axialvector-axialvector-type  and vector-vector type scalar tetraquark currents and scalar-tensor type tensor tetraquark current to study the scalar, vector and axialvector tetraquark states with
the QCD sum rules by carrying out the operator product expansion up to vacuum condensates of dimension $10$ in a consistent way. In calculation, we use the energy scale formula to determine the ideal energy scales of the QCD spectral densities to extract the masses from the QCD sum rules with the pole contributions about $(40-60)\%$. The present calculations do not favor assigning the $Z_c(4100)$ to be the scalar or vector tetraquark state. If the $Z_c(4100)$ is a scalar tetraquark state without mixing effects, it should have a mass about $3.9\,\rm{GeV}$ or $4.0\,\rm{GeV}$ rather than $4.1\,\rm{GeV}$;
on the other hand, if the $Z_c(4100)$ is a vector tetraquark state, it should have a mass about  $4.2\,\rm{GeV}$ rather than $4.1\,\rm{GeV}$.
If we introduce mixing effects,  a  $\tilde{A}\tilde{A}\oplus\tilde{V}\tilde{V}$  type mixing scalar tetraquark state can  have a mass about $4.1\,\rm{GeV}$.
More precise measurements  of the mass, width and quantum numbers are still needed.
 As a byproduct, we obtain an axialvector tetraquark candidate for the $Z_c(4020)$.

\section*{Acknowledgements}
This  work is supported by National Natural Science Foundation, Grant Number  11775079.

\end{document}